\documentclass[11pt]{article}
\usepackage{color}

\usepackage{latexsym}  
\usepackage{amssymb}
\usepackage{graphicx}
\usepackage{amsmath}
\usepackage{mathrsfs}

\newcommand{\be}{\begin{equation}}
\newcommand{\ee}{\end{equation}}
\newcommand{\bea}{\begin{eqnarray}}
\newcommand{\eea}{\end{eqnarray}}
\newcommand{\bref}[1]{(\ref{#1})}

\newcommand{\pa}{\partial}

\begin{document}
\begin{titlepage}
\begin{flushright}
\today
\end{flushright}

\begin{center}
{\Large\bf  
Ultralight axion and modern cosmology tensions}
\end{center}

\begin{center}

\vspace{0.1cm}

{\large Takeshi Fukuyama$^{a,}$%
\footnote{E-mail: fukuyama@rcnp.osaka-u.ac.jp}}

\vspace{0.2cm}

{\small \it ${}^a$Research Center for Nuclear Physics (RCNP),
Osaka University, \\Ibaraki, Osaka, 567-0047, Japan}



\end{center}
\begin{abstract}
String-inspired axion model is considered to comprehensively solve the problems of modern cosmology, Hubble tension problem,  the origin of the stochastic gravitational wave background, and too early formation of supermassive black holes at high $z$.

\end{abstract}
\end{titlepage}
\section{Introduction}
There are many complicated unsolved problems in modern cosmology. 
They are the origins of dark matter (DM) and dark energy (DE),  Hubble tension problem \cite{Aghanim, Riess, Valentino},  and the origin of stochastic gravitational wave background (SGWB) found by the North American Nanoherz Observatory for Gravitational Waves (NANOGrav) \cite{Nano1, Nano2} and Pulsar Timing Array Collaborations (PTAs) \cite{Antoniadis, Reardon} etc.
 In the recent letter \cite{Fukuyama} we showed that the string-inspired axion motivated by the strong CP problem in particle physics \cite{PQ} solves the problem why primordial supermassive massive black holes (SMBHs) appeared so early in the Universe \cite{Richstone1998, Kormendy, Banados, JWST} and briefly disccussed the possibility that it also causes the SGWB. 
 In this letter, we show that this ultralight axion with mass around $O(10^{-23})$ eV can also solve comprehensively all the above mentioned problems. 
 In Sec.2 we review the string-inspired axion briefly. In Sec.3 we explain how it saves the problems mentioned above. 
 
 We adopt $\hbar=c=1$ units in this letter.
 \section{Review of string-inspired axion}
 The string-inspired axion is the axion model whose breaking scale is extended to string region \cite{Soda2020, Witten, Arvanitaki, Visinelli}.  The original QCD axion potential based on the chiral symmetry breaking and $\theta$ vacuum 
 is given by
\be
V(\phi)=m_u\Lambda_{QCD}^3e^{-S_{instanton}}\left[1-\cos\left(\frac{\phi}{f_a}\right)\right]\equiv \Lambda^4\left[1-\cos\left(\frac{\phi}{f_a}\right)\right],
\label{potential}
\ee
where
\be
S=\frac{1}{4g^2}\int d^4x Tr[F_{\mu\nu}F^{\mu\nu}]
\ee
with $F_{\mu\nu}\equiv F_{\mu\nu}^a\frac{\lambda^a}{2}$.
Instanton of winding number=1 is $\tilde{F}_{\mu\nu}\equiv \frac{1}{2}\epsilon_{\mu\nu\rho\sigma}F^{\rho\sigma}=F_{\mu\nu}$ and
\be
S_{instanton}=\frac{1}{4g^2}\int d^4x Tr[F_{\mu\nu}\tilde{F}^{\mu\nu}]=\frac{8\pi^2}{g^2}.
\ee
So in the strong coupling regime of $g$, $e^{-S_{instanton}}$ contribution is mild.
Using the Gellmann-Oakes-Renner relation \cite{Gellmann}, 
\be
\Lambda_{QCD}^3=\frac{F_\pi^2m_\pi^2}{m_u+m_d} 
\label{lambda}
\ee
Here $F_\pi=93$ MeV, and $\frac{m_u}{m_d}\approx 0.47$, and we obtain axion mass
\be
m_a=5.7\times 10^{-6}\left(\frac{10^{12}\mbox{GeV}}{f_a}\right) \text{eV}.
\label{ma1}
\ee
Originally, in order to obtain SMBH at high $z$, we were forced to make $m_a$ much smaller than \bref{ma1} \cite{Fukuyama} and to proceed to higher symmetry breaking scale, the string-inspired axion.
\be
V(\phi)=\Lambda_{string}^4\left[1-\cos\left(\frac{\phi}{f_a}\right)\right],
\label{potential3}
\ee
where
\be
\Lambda_{string}^4=M_{SUSY}^2M_{Pl}^{*2}e^{-S_{instanton}}
\label{SUSY}
\ee
with $M_{Pl}^*=\sqrt{\frac{1}{8\pi G}}=2.4\times 10^{18}$ GeV.
So $f_a$ is increased, that is, if the energy scale $\mu$ is increased, the running strong coupling constant $g(\mu)$ is reduced, and we have ultralight axion mass from \bref{SUSY}.
Here $f_a$ and axion mass $m_a$ become independent parameters unlike QCD axion and are given as follow:  $f_a$ is \cite{Witten}
\be
f_a=\frac{\alpha_{GUT}M_{Pl}^*}{\sqrt{2}2\pi}\approx 1.1\times 10^{16}~\mbox{GeV}.
\label{GUT}
\ee
Here we have set the strong coupling constant $\alpha_{GUT}$ at GUT scale as
\be
\alpha_{GUT}\equiv \frac{g_{GUT}^2}{4\pi}=\frac{1}{25}.
\ee
and naively
\be
\frac{S_{instanton}}{2}=\frac{\pi}{\alpha_{GUT}}\approx 79.
\ee
However, there are many parameters in axiverse \cite{Arvanitaki} like $M_{SUSY},~S_{instanton},~\mbox{misalignment parameter}~\theta_i,$
primordial isocurvature fraction. Visinelli and Vagnozzi obtained 
\be
S_{instanton}=198\pm 28~~ \mbox{and} ~~M_{SUSY}=10^{11} \mbox{GeV}
\label{instanton}
\ee
by Bayesian parameter inference in light of many cosmological data \cite{Visinelli} \footnote{This SUSY breaking scale seems to be too large in comparison with that obtained from the gauge coupling unification \cite{Fuku}. }. 
Then, if we adopt the center value $198$ for $S_{instanton}$, we obtain
\be
\mbox{log}_{10}(m_a/\mbox{eV})=-21.5^{+1.3}_{-2.3}.
\label{ma2}
\ee
Then we proceed to discuss that this ultralight axion solves the above mentioned cosmological problems.
\section{The string-inspired axion solves the cosmological problems}
\subsection{Hubble tension problem}
Firstly we discuss on the Hubble constant problem which implies the tension between the early $H_0(\mbox{CMB})=67 \mbox{Km}~ \mbox{s}^{-1}~\mbox{Mpc}^{-1}$ from CMB and the late $H_0(\mbox{SNe}) =73 \mbox{Km}~ \mbox{s}^{-1}~\mbox{Mpc}^{-1}$ from SNe Ia. 
As we have shown in \cite{Fukuyama}, axion behaves as DE when $m_a t\ll 1$ and does as DM when $m_at\gg 1$ \footnote{In this calculation, we have neglected the self-interaction term, which is safely satisfied in ultralight axion mass.}.
Thus axion begins to oscillate at the scale factor $a_{osc}(t_{osc})$ with $m_at_{osc}=1$ satisfying
\be
a_{osc}=\left(\frac{t_{osc}}{t_{eq}}\right)^{1/2}\left(\frac{t_{eq}}{t_0}\right)^{2/3}a_0=\sqrt{\frac{H_0}{m_a}}a_{eq}^{1/4}
\ee
Substituting $m_a=10^{-23}$ eV and $H_0=10^{-33}$ eV, we obtain $a_{osc}=10^{-6}$.
See Fig. 1.
\begin{figure}
\begin{center}
 \includegraphics[width=6cm]{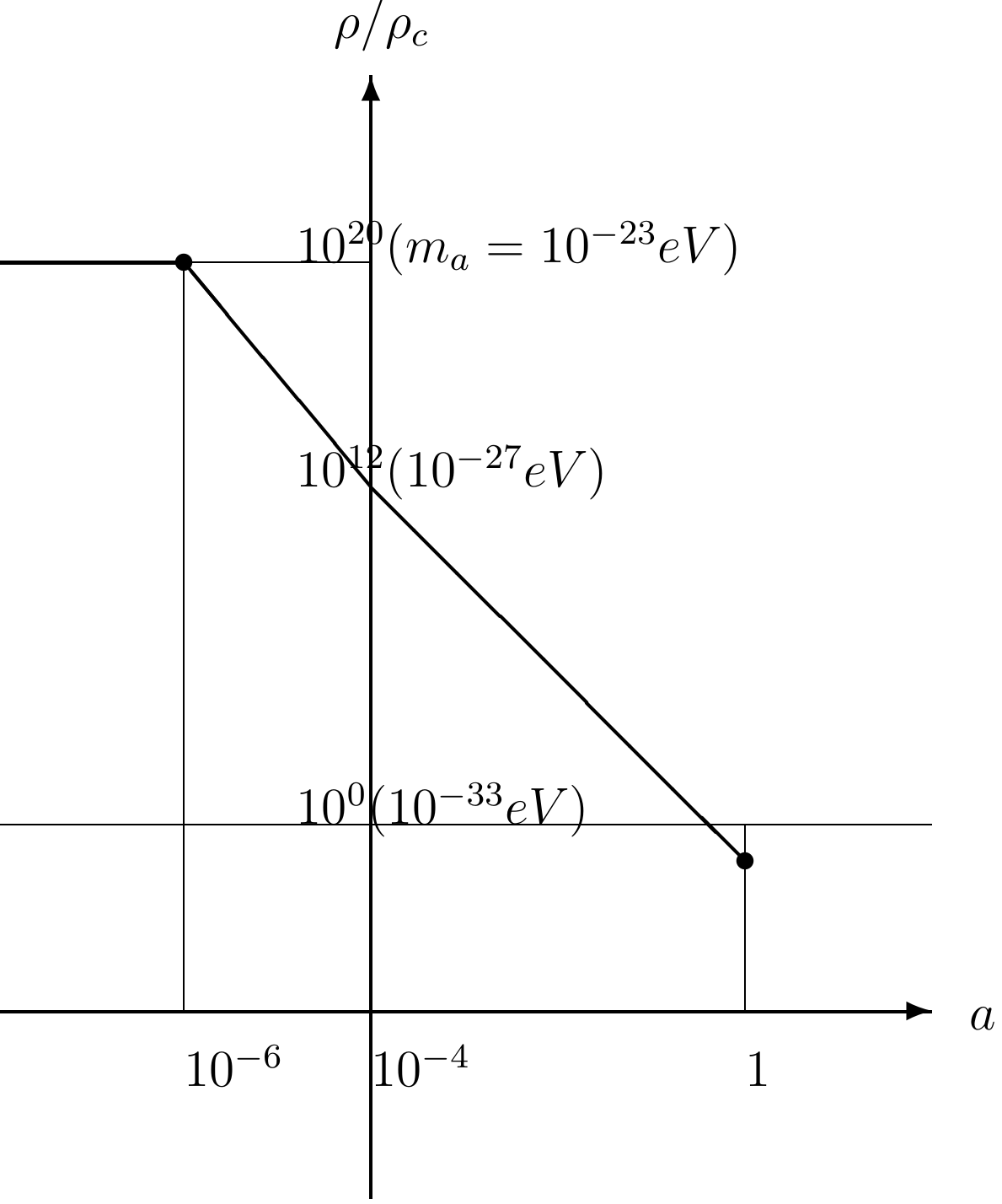}
\caption{ The evolution of DM and DE (bold line): Axion behaves as DE at $a<a_{osc}=10^{-6}$, and starts to oscillate at $a_{osc}$ in radiation dominant era ($\rho\propto a^{-4})$, and at $a=a_{eq}=10^{-4}$ it enters into matter dominant era ($\rho\propto a^{-3}$), reaching to the $\Omega_0\approx 0.3$ at $a=a_0=1$.}
\end{center}
\end{figure}
Here the Hubble constant satisfies the Friedmann equation,
\be
H(t)^2=\frac{8\pi G}{3}\rho(t)+\frac{\Lambda}{3}-\frac{K}{a(t)^2}
\ee
with the assumption $K=0$.
The Hubble constant from CMB observation is determined from the angular diameter distance 
\be
D_A=\frac{r_s}{\theta^*}\propto \frac{1}{H_0}.
\label{DA}
\ee
Here the $r_s$ is the sound horizon defined by
\be
r_s\equiv \int_{z^*}^\infty \frac{c_s}{H(z)}dz,
\label{rs1}
\ee
Here $z^*$ is the redshift of the recombination at $3000$ K and roughly of the same order as  $z_{eq}$ at $9000$ K.

 $c_s$ is the sound velocity defined by
\be
c_s=\frac{1}{\sqrt{3(1+R(z))}}=\frac{k}{2m_a},
\ee
where $R(z)$ is the baryon-loading factor
\be
R=\frac{3\Omega_B}{4\Omega_\gamma}\frac{a}{a_0}
\ee
The second equality comes from the quantum pressure $P$ defined by
\be
P=-\frac{1}{4m_a^2a^2}\Delta \delta=\frac{k^2}{4m_a^2a^2}\delta,
\ee
with $\delta\equiv \frac{\delta \rho}{\rho_{background}}$,
and
\be
c_s^2=\rho\frac{\delta P}{\delta\rho}=\frac{k^2}{4m_a^2a^2}.
\ee

If the universe has the additional DE due to axion degeneracy mentioned above,  the sound vanishes at $a_{osc}$ since $\phi(t)$=const. at $m_at<1$ as mentioned above and $r_s$ is modified as
\be
r_s\equiv \int_{z^*}^{z_{osc}} \frac{c_s}{H(z)}dz.
\label{rs2}
\ee
That is, $H_0$(CMB) calculated from \bref{rs1} is underetimated compared with the true one (which should be equal to $H_0(\mbox{SNe})$) estimated from \bref{rs2} by the factor
\be
\frac{H_0(\mbox{SNe})}{H_0(\mbox{CMB})}=\frac{1/(1+z^*)}{1/(1+z^*)-(1/(1+z_{osc})}\approx 1+\frac{z^*}{z_{osc}}.
\ee
For $m_a=10^{-23}$ eV, we obtained $z_{osc}=10^6$ and $z^*=10^4$, and the discrepancy between $H_0$(CMB) and $H_0$(SNe) is withing good coincidence.
We know that there are arguments on ``axion like'' models which insist axion potential 
\be
V(\phi)\propto \left[1-\cos\left(\frac{\phi}{f_a}\right)\right]^n~~(n>1)
\ee
for more detailed adjustment. 
However, we want to adhere to \bref{potential} having the definite background on the heterotic string theory \cite{Soda2020, Witten}.

Next, we show that the coherent oscillation of axion DM induces the observed SGWB \cite{Rubakov}.
\subsection{The origin of the stochastic gravitational wave background}
We show that the coherent oscillation of axion DM whose mass is around $10^{-23}$ eV induces the observed SGWB.

Indeed, the metric in the Einstein equation is generalized from the Schwarzshild form to
\be
ds^2=(1+2\Phi({\bf x},t))dt^2-(1+2\Psi({\bf x},t))\delta_{ij}dx^idx^j,
\label{Schwarz}
\ee
and $\Phi$ and $\Psi$ are divided into the static and oscillation parts \cite{Rubakov},
\bea
\Phi({\bf x},t)&=&\Phi_0({\bf x})+\Phi_c({\bf x})\cos (\omega t+2\alpha({\bf x}))+\Phi_s({\bf x})\sin  (\omega  t+2\alpha({\bf x})),\\
\Psi({\bf x},t)&=&\Psi_0({\bf x})+\Psi_c({\bf x})\cos (\omega t+2\alpha({\bf x}))+\Psi_s({\bf x})\sin  (\omega  t+2\alpha({\bf x}))
\eea
with
\be
\Delta\Phi_0=4\pi G\rho_{DM} ~\mbox{and}~\Phi_0=-\Psi_0
\ee
etc.
The oscillation part satisfies 
\be
6\ddot{\Psi}-2\Delta (2\Psi+\Phi)=8\pi G\rho_{DM}(1+3\cos 2m_at).
\label{Ein1}
\ee
Here the background satisfies
\be
\Delta\Phi_0=4\pi G\rho_{DM},~~\Phi_0=-\Psi_0.
\ee
From \bref{Ein1}, we obtain the oscillation
\be
\delta \Psi=-\frac{\pi G\rho_{DM}}{m_a^2}\cos2m_at
\ee
and SGWB amplitude $h_c$ is
\be
h_c=2\sqrt{3}\delta \Psi=2\times 10^{-15}\left(\frac{\rho_{DM}}{0.3\mbox{GeV}/cm^3}\right)\left(\frac{10^{-23}\mbox{eV}}{m_a}\right),
\ee
and the frequency $f$,
\be
f=5\times 10^{-9}~\mbox{Hz}\left(\frac{m_a}{10^{-23}\mbox{eV}}\right).
\ee
This may be within 2$\sigma$ of observations \cite{Nano1}. We need further measurements \cite{SKA} towards more definitive conclusions.

Lastly, we show briefly that this axion saves the too early formation of supermassive BH. Please reffer \cite{Fukuyama} for the details.
\subsection{Too early formation of supermassive BH at high $z$.}
The accretion processes of both baryonic and collisionless DM are rather slow. So it is very difficult to explain why SMBHs have been observed at high $z$. In this subsection we explain the self attractive interaction of axion of \bref{potential3} saves this problem.
So far we have discussed the process coupled with cosmological expansion. In this section we discuss the local process decoupled with expansion. Corresponding to the situation, 
the axion is divided into fast oscillation ($e^{im_at}$) part and slow one ($\psi$) as
\be
\phi=\frac{1}{\sqrt{2m_a}}\left(\psi e^{-im_at}+\psi^*e^{im_at}\right).
\label{Eq5}
\ee
Taking account of the axion self coupling \bref{potential3}, the axion Lagrangian density for $\psi$ becomes
 \be
\mathcal{L}= \frac{i}{2}\left(\psi^*\frac{\pa\psi}{\pa t}-\psi\frac{\pa\psi^*}{\pa t}\right)-\frac{1}{2m_a}\nabla\psi^*\cdot\nabla\psi-\frac{g_dN}{2}|\psi|^4+N|\psi(x)|^2\int\frac{Gm_a^2}{|{\bf x}-{\bf y}|}|\psi(y)|^2d^3y.
\ee
Here $N$ is the total number of axion particles in the overdense region. Dimensional quadratic coupling $g_d$ is related with the self-coupling $\lambda$ with $\frac{\lambda}{4!}\phi^4$ term in \bref{potential3} by
\be
g_d=\frac{\lambda}{m_a^2}=\frac{4\pi a_s}{m_a}<0,
\label{g}
\ee
where $|a_s|$ is the scattering length. Here we consider the fluctuation (the overdense region) of the axion field decoupled from the cosmological expansion and use Gaussian approximation \cite{Gupta2017} with angular momentum $l$, 
\begin{equation}
|\psi\left(t,x\right)|=\frac{1}{\sqrt{2\pi(l+1)!\sigma^3}}\left(\frac{r}{\sigma}\right)^le^{-r^2/(2\sigma^2)}Y_{lm}(\theta, \phi).
\end{equation}
Then we obtain the effective potential,
\be
V_{eff}=\frac{1}{2m_a\sigma^2}-\frac{\sqrt{2}}{3\pi}\frac{GNm_a^2}{\sigma}+\frac{l(l+1)}{2m_a\sigma^2}+\frac{g_dN}{6\sqrt{2}\pi^{3/2}\sigma^3}.
\label{potential2}
\ee

The effective potential \bref{potential2} has the stable orbits at \cite{Eby, Hertzberg}
\be
\sigma_{min}=\frac{\frac{\tilde{l}^2}{2m_a}\pm \sqrt{\left(\frac{\tilde{l}^2}{2m_a}\right)^2+\frac{g_dGN^2m_a^2}{6\pi^{5/2}}}}
{\frac{\sqrt{2}}{3\pi}GNm_a^2} 
\label{min}
\ee
with 
\be
\tilde{l}^2\equiv l(l+1)+1.
\ee
\bref{min} allows the two extrema in general, and the minus (plus) case corresponds to stable (unstable) orbit. 
Here it is very important that axion DM attracts to each others since $g_d<0$. These maximum and minimum points coalesce at $\sqrt{{...}}=0$ in \bref{min} and the stable orbit disappears, leading to so called 
Kaup radius,
\be
\sigma_{Kaup}=\frac{\sqrt{3}}{2\pi^{1/4}}\frac{\sqrt{-g_d}}{\sqrt{G}m_a}=\sqrt{-6\pi^{1/2}g_d}\frac{M_{Pl}^*}{m_a}
\label{Kaup}
\ee
with the Kaup mass,
\be
M_{Kaup}=Nm_a=\frac{\sqrt{6\pi^{5/2}}}{2}\frac{\tilde{l}^2}{\sqrt{G|\lambda |}}
=25\tilde{l}^2\frac{M_{Pl}^*}{\sqrt{|\lambda|}}.
\label{mass}
\ee
This mass is eventually reduced to BH mass since there is no repulsive force to prevent the collapse,
\be
M_{Kaup}=M_{BH}.
\ee
Using the values of \bref{instanton} and \bref{ma2}, we obtain 
\be
M_{BH}=2.5~\tilde{l}^2\left(\frac{10^{-21}\mbox{eV}}{m_a}\right)\times 10^8M_\odot .
\label{SMBH}
\ee

\section{Conclusion}
We have shown that the ultralight axion originally came from the strong CP violation problem saves the miscellaneous cosmological problems like Hubble tension problem, the origin of SGWB.  Moreover, even after the axion decoupled with cosmological expansion, its self-attraction causes the early formation of SMBH.
Thus, string-inspired axion comprehensively solves the miscellaneous problems ranging from the Planck scale to the cosmological scale. 

\section*{Acknowledgments}
We would express our sincere thanks to Dr. J. Soda for useful discussions and comments. We are also grateful to Dr. K. Shigemoto for useful discussions.
This work is supported by JSPS KAKENHI Grant Numbers JP22H01237.

\end{document}